# Influence of Plasma Density Arrangement on Millimeter-Wave Transmission Characteristics

Wenbo Liu[1], Peian Li[1], Guohao Liu[1,2], Jianjun Ma[1,2,3], Houjun Sun[1,2]

[1]School of Integrated Circuits and Electronics, Beijing Institute of Technology, Beijing 100081, China;
[2]Tangshan Research Institute, BIT, Hebei, 063099 China
[3]Beijing Institute of Technology Yangtze River Delta Research Institute (Jiaxing), Zhejiang, China
E-mail: jianjun_ma@bit.edu.cn

*Abstract*—The advancement of millimeter-wave and terahertz technologies have revolutionized high-speed wireless networks and precise tracking systems. These technologies offer unique penetration capabilities in specific scenarios, significantly enhancing the capability to investigation plasma. Recent breakthroughs include the precise diagnosis of plasma electron density using terahertz-time domain spectroscopy and the modeling of plasma sheaths in re-entry spacecraft through scattering matrices. Concurrently, extensive research efforts have been dedicated to comprehending plasma's influence on electromagnetic wave behaviors, encompassing reflection, transmission, absorption and also phase shift. In this paper, we employ COMSOL Multiphysics software to create an inductively coupled plasma (ICP) device, enabling the simulation of various plasma density arrangements. Our investigation focuses on unraveling the intricate interplay between plasma configurations and millimeter-wave transmission characteristics. The findings underscore the substantial impact of diverse plasma concentration arrangements on the behavior of electromagnetic waves traversing through them. Additionally, these arrangements endow the plasma with a discernible degree of frequency selectivity, thus expanding our understanding of plasma behavior in novel ways.

*Keywords—Millimeter wave, inductively coupled plasma; plasma arrangement; transmission*

## I. INTRODUCTION

Advancements in communication technologies, coupled with the progress in millimeter-wave and terahertz technologies, have opened doors to high-speed, high-capacity wireless networks and high-precision tracking and detection systems [1]. Simultaneously, these technological developments have greatly facilitated research in the realm of plasma, thanks to the remarkable penetration capabilities of millimeter waves in specific scenarios[2]. Recent breakthroughs in terahertz time-domain spectroscopy (THz-TDS) have enabled precise diagnosis of plasma electron density[3, 4], yielding more accurate insights into electron density distribution and generation mechanism within inductively coupled plasma (ICP)[5].

Moreover, the electromagnetic properties of plasma have become a focal point in the domain of electromagnetic wave interference. This is particularly relevant for applications like communication and detection systems for re-entry spacecraft. Researchers have used flow field models to establish plasma parameter distributions and employed theoretical calculations, such as scattering matrices, to evaluate the transmission characteristics of terahertz waves through plasma sheath[6, 7]. Concurrently, efforts to comprehend the impact of plasma on electromagnetic wave phenomena, including reflection, transmission, absorption, and phase shift, are ongoing[8-10]. Notably, a study has hinted at the promising potential of leveraging plasma's electromagnetic properties, especially in the realm of radar stealth technology[11].

In the next generation of communication and radar equipment, the millimeter-wave frequency band, particularly in the terahertz range, is widely acknowledged as the future frontier[12]. Despite the well-documented penetrative capabilities of millimeter waves, the influence of varying plasma density arrangements on millimeter-wave transmission characteristics has not received adequate attention, primarily due to prevailing hardware limitations. This paper addresses this gap by employing COMSOL Multiphysics software to construct an inductively coupled plasma (ICP) device. Through simulations, we explore two distinct plasma density arrangements and conduct a comprehensive analysis of millimeter-wave transmission properties.

## II. SIMULATION SYSTEM AND SIMULATION DESIGN

### A. Description of the inductively coupled plasma source

In this work, we harnessed the capabilities of 'COMSOL Multiphysics' software to conceive and fabricate an Inductively Coupled Plasma (ICP) device. The ICP plasma source, illustrated in Figure 1, is composed of two primary components: an RF antenna and a hermetically sealed quartz tube, serving as the essential elements responsible for generating and confining a quasi-neutral plasma.

The RF antenna consists of an ideal solid copper coil, boasting a diameter of 5mm. For the purposes of our analysis, we do not take into account thermal decay and thermal losses. It is wound around a cylindrical quartz tube with a thickness of 2mm, which is 200mm long and has a diameter of 100mm, forming 9 turns. The RF power source operates at 150W with a common excitation frequency of 13.56MHz[13], and the induction region within the quartz tube, filled with argon gas at a pressure of 1 torr to generates an RF glow discharge plasma.

Throughout the simulation process, the ICP system reaches a state of equilibrium and stability in 1 second after the reaction is initiated. It can maintain a plasma with electron density ranging from $2.95 \times 10^{19}$ to $2.75 \times 10^{14}$. The electron density distribution is shown in Fig. 1, displaying a decreasing trend from the center towards the outer regions.

*B. Plasma arrangement & Millimeter-wave transmission*

To eliminate the lens effect caused by the non-uniform longitudinal distribution of plasma[2], we employed narrow-beam millimeter waves with a beam width of 1-2 cm for transmission simulations, covering a frequency range from 10 GHz to 300 GHz. This approach allowed us to treat the beam coverage area as a uniform layered structure, as depicted in Fig. 2. Leveraging the electron density distribution within the ICP and the mode of millimeter wave propagation, our simulation involved finite element analysis. We selected the beam's propagation center axis as the data sampling line and reconstructed the electron density in layers (comprising a total of 22 layers, including air layers). The outermost layers flanking the reaction vessel consist of air layers with a refractive index of 1. As the electromagnetic wave penetrates deeper into the reaction vessel's core, the electron density gradually increases, forming a concentration distribution pattern of 'low-high-low'.

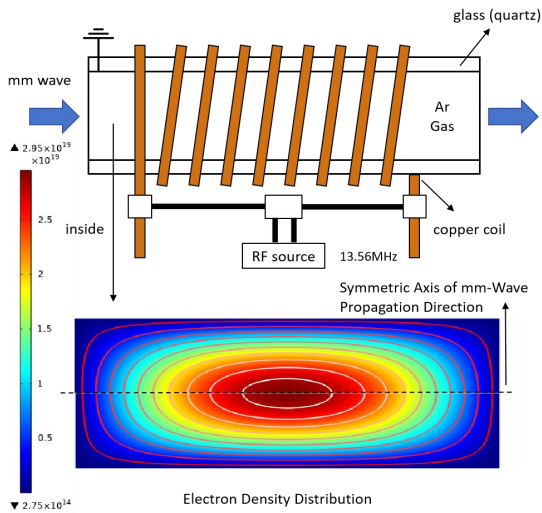

Fig. 1. A schematic & Electron Density Distribution of the inductively coupled plasma source for the simulation.

Concurrently, we utilized ICP simulation data to establish an alternative plasma arrangement. As depicted in Fig. 2, millimeter wave transmission followed an electron density variation path encompassing 'air - high concentration-low concentration - high concentration - air.' This setup allowed us to examine the influence of arrangement on millimeter wave transmission under equivalent concentration conditions.

## III. RESULT AND ANALYSE

We conduct independent frequency sweep transmission simulations covering a wide range from 10 GHz to 300 GHz for two distinct plasma density arrangements. Notably, both arrangements exhibited exceptional frequency cutoff characteristics when the transmission frequency exceeded the plasma oscillation frequency (cutoff frequency) of 48 GHz.

When the incident wave frequency fell below the cutoff frequency, the beam encountered a total reflection state upon interacting with the plasma layer. Conversely, when the incident electromagnetic wave's frequency aligned with the plasma oscillation frequency, it experienced maximum absorption effects. This absorption effect gradually diminished as the incident frequency increased, leading to the observed variations depicted in Figures 3a, b, and c.

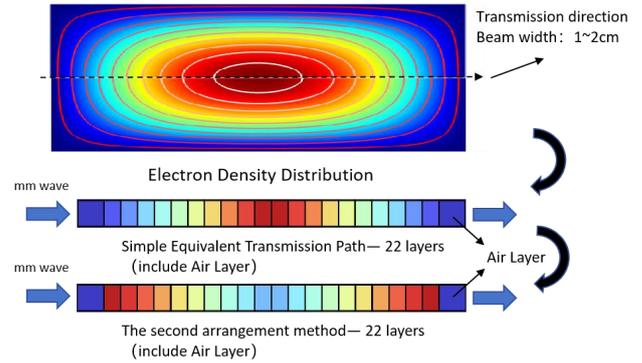

Fig. 2. Electron Density Distribution of the ICP source for the simulation, and the Plasma arrangement.

For incident waves propagating through the plasma with the 'low-high-low' concentration configuration, an intriguing trend emerged. As the incident frequency increased, the wave penetrated an increasing number of plasma layers, resulting in heightened absorptivity and reduced reflectivity, particularly within the frequency range of 25 GHz to 48 GHz. In contrast, the 'high-low-high' arrangement within the 48 GHz to 100 GHz frequency range exhibited characteristics akin to a grating, showcasing distinct reflection peaks at specific frequencies, such as 53 GHz, 62 GHz, 67 GHz, and 71 GHz.

Facilitating the observation of phase difference characteristics, Fig. 3d depict data offset by $3\pi$ radians. It becomes evident that both arrangement modes—'high-low-high' and 'low-high-low'—yield nearly identical transmission phase shifts. Intriguingly, the arrangement patterns exhibit minimal impact on the transmission phase. It is noteworthy that the theoretical calculation curves employ average density for phase shift calculations, introducing disparities between the theoretical and simulated data within the 48GHz to 100GHz frequency range.

As the frequency approaches the cutoff frequency, variations in plasma permittivity become more pronounced. Complex multi-layer structures with varying concentrations resist effective simplification into a single averaged concentration structure. Consequently, notable disparities between calculated and simulated data curves arise. However, in frequency ranges characterized by minimal differences in permittivity, the simulated curves closely align with the numerical calculation curves, particularly above 100GHz.

## IV. CONCLUSION

Varying the concentration arrangements of plasma yields distinct electromagnetic wave penetration and transmission behaviors, particularly evident when the wave frequency approaches the plasma frequency. Manipulating plasma concentration and arrangement offers a promising avenue for achieving diverse frequency absorption and selection characteristics. Effectively harnessing this capability may prove pivotal in advancing plasma stealth and filtering technologies.

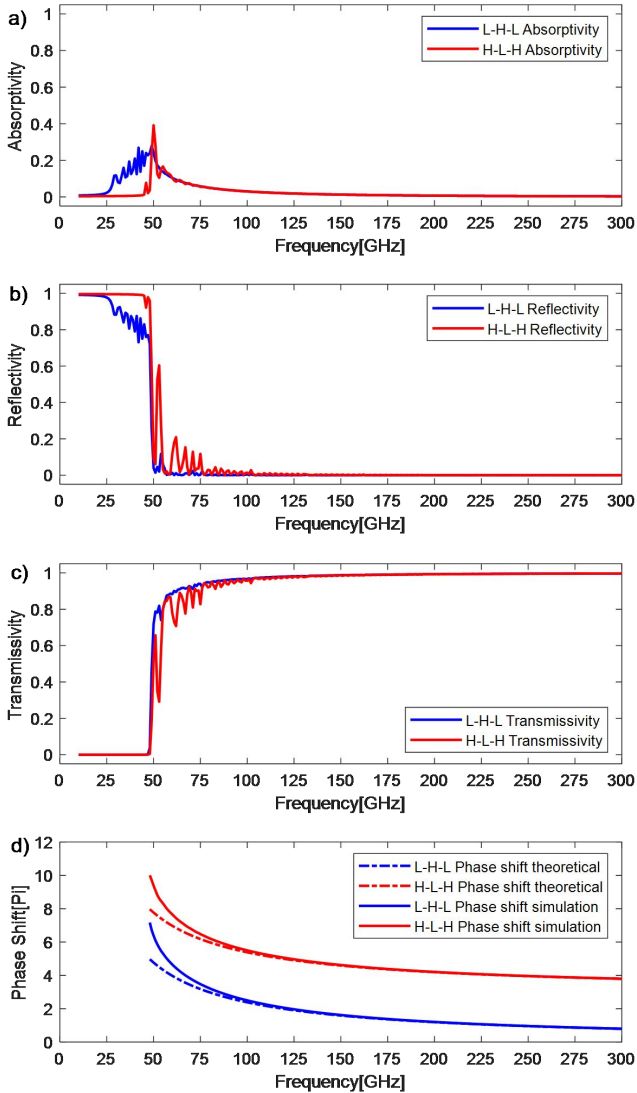

Fig. 3. Frequency sweep simulation experiment of the "high-low-high" and "low-high-low" concentration arrangements. a) Absorptivity of two concentration arrangements. b) Reflectivity of two concentration arrangements. c) Transmissivity of two concentration arrangements. d) Simulation phase shift & theoretical calculated phase shift between two concentration arrangements.

The findings highlight that in scenarios featuring a ′high-low-high′ concentration arrangement, electromagnetic waves with frequencies exceeding the cut-off frequency (48GHz-75GHz) exhibit significant absorption and multiple reflection peaks. Conversely, for the 'low-high-low' concentration arrangement, frequencies lower than the cutoff frequency (25 GHz to 48 GHz) display relatively low reflectivity and high absorption rates. Importantly, both concentration arrangements do not introduce substantial interference to signal phase, and with increasing frequency, the transmission phase difference closely aligns with theoretical expectations.

In light of these findings, we emphasize the non-negligible impact of varying plasma concentration arrangements and distributions on electromagnetic waves transmitted through them. Further research in related fields should delve deeper into unlocking advancements in technologies like plasma barriers and stealth. While this work explored numerous plasma arrangement configurations and millimeter-wave penetration transmission studies, not all results are presented herein.

ACKNOWLEDGMENT

This work was supported in part by the National Natural Science Foundation of China (62071046), the Science and Technology Innovation Program of Beijing Institute of Technology (2022CX01023), the Graduate Innovative Practice Project of Tangshan Research Institute, BIT (TSDZXX202201)and the Talent Support Program of Beijing Institute of Technology "Special Young Scholars" (3050011182153)